\RequirePackage{fix-cm}
\documentclass[natbib,smallextended]{svjour3}       
\smartqed  
\usepackage{graphicx}
\usepackage{aps-bibstyle}  
\usepackage[extd]{journalnames}
\usepackage{amsmath}
\usepackage{amssymb}
\usepackage{lscape}
\usepackage[bookmarks=true, bookmarksopen=true]{hyperref}
\journalname{Space Science Reviews}
\begin{document}

\title{Insights into planet formation from debris disks: I.  The solar system as an archetype for planetesimal evolution
}
\label{s:birth}



\author{ Brenda C. Matthews \and       
 JJ Kavelaars 
}

\authorrunning{Matthews \& Kavelaars} 

\institute{
           B. C. Matthews \and J. J. Kavelaars \at
              National Research Council of Canada, 5071 West Saanich Road,
              Victoria, BC, V9E 2E7 Canada \\
              University of Victoria, Finnerty Road,
              Victoria, BC V8W 3P6, Canada \\
}

\date{Received: date / Accepted: date}

\maketitle


\abstract{Circumstellar disks have long been regarded as 
windows into planetary systems. The advent of high sensitivity, high
resolution imaging in the submillimetre where both the solid and gas
components of disks can be detected opens up new possibilities for
understanding the dynamical histories of these systems and therefore,
a better ability to place our own solar system, which hosts a
highly evolved debris disk, in context. Comparisons of dust masses
from protoplanetary and debris disks have revealed a stark downturn in
mass in millimetre-sized grains around a stellar age of 10 Myr,
ostensibly in the ``transition disk'' phase, suggesting a period of
rapid accretion of such grains onto planetesimals. This rapid
formation phase is in keeping with radionucleide studies of Kuiper
Belt Objects in the solar system. Importantly, this suggests that any
thermal gradients in the gas of disks of this era will be ``frozen
in'' to the planetesimals as they rapidly accrete from the solids and
ices in their vicinity.  Measurements of radial gradients in thermal
tracers such as DHO, DCN and other tracers can therefore provide
insight into the nascent solar system's abudances. In studies of
dynamical evolution of the solar system, it is tacitly assumed that
such abundances can reveal the location of formation for bodies now
found in the asteroid belt and Kuiper belt. Similarly, evidence of gas
detected from collisional evolution in young debris disks could
potentially reveal how rapidly objects have dynamically evolved in
those systems, most of which will be significantly younger than the
solar system. }


\section{Introduction}

Circumstellar disks are found around stars of all ages. 
Around forming stars, protostellar disks play a critical
role in the accretion of material onto forming stars, and over time,
the disks can evolve to protoplanetary disks, wherein stable
locations in the midplane allows agglomeration processes to assemble
the solids from sub-micron sized dust to grains up to centimetres in
size, crucial steps in the planet formation process 
\citep{2008ARA&A..46...21B,2014prpl.conf..339T}.

As protoplanetary disks evolve, they become more tenuous,
eventually beginning to erode from the inside out, creating disks with
inner holes, coinciding with a decline in infrared excess emission
such that half the disks are gone by 3 Myr and no disks are detectable
by 6 Myr, based on a study of young clusters by
\cite{2001ApJ...553L.153H}.  During the transition disk epoch, gas can
continue to accrete onto planetary embryos and dust grains can
continue to grow. Ultimately however, in the late stages of a
transition disk, the remaining dust and gas assembled from the natal
star-forming cloud or core is dispersed
\citep{espaillat_etal_2014}.  The decline of near-IR emission,
the loss of gas components of the disks and the changes in the dust
grain size distribution are all identifiers of transitional objects,
but it is not immediately clear whether these changes occur
simultaneously or in a sequence
\citep[e.g.,][]{2015Ap&SS.357..103W}.  Therefore, at the end of the 
transition disk stage, the formation of gas giants and super Earths
must be complete, although terrestrial planet formation may continue
for up to 100 Myr (see the Chassefi\`ere chapter in this work).
After the assembly of planetesimals and planets is complete, some
parts of the disk may already be dominated by destructive collisions,
rather than agglomerative processes, although it is likely destructive
collisions occur even from the earliest agglomeration stages in
gas-rich disks
\citep{2005A&A...434..971D}. When a disk is predominately populated by
dust and gas produced in collisions, it is known as a debris disk.
Historically, such disks were also referred to as ``secondary'' or
``second-generation'' disks owing to their content being entirely
derived from collisional evolution of larger bodies, rather than any
remnants of the initial protoplanetary disk. 

Debris disks represent the longest lived phase of circumstellar disks,
ranging from the youngest systems, many of which now are seen to still
host remnant protoplanetary gas disks, to very evolved systems,
including sub-giant stars
\citep{2014MNRAS.437.3288B,2013MNRAS.431.3025B} and white dwarfs, where
the measured incidence rates of infrared excess are typically a few
percent \citep[e.g.,][]{barber12}, but have been measured to be as high as 14\%
\citep{2007ApJ...660..641K}. White dwarfs also show evidence of ``pollution'' of their atmospheres due to deposition of
material, most likely from a circumstellar disk or remnant
planetesimals \citep[e.g.,][]{zuckerman03,dufour12}. All these evolved
systems have in common the fact that their protoplanetary disk phase
and succeeded in forming planetesimals of at least 100 km in size.
Such objects may be essential for planetary assembly, but not all find
their way into planets.  Instead, the collisions of these
planetesimals provide the source of debris disk dust emission for the
lifetime of the system.

The sun is a middle-aged star, and we see its debris disk components
primarily through direct detection of the planetesimals (asteroids and
comets) that make up the largest end of its size distribution. 
Currently, collisions within the Kuiper Belt occur on very long
timescales, meaning its mass is relatively constant, but its dust
emission is relatively low.  Breakup of large objects can still occur,
however, as evidenced by the asteroid P/2010 A2, which is observed to
be fragmenting, either due to impact or rotationally induced breakup
\citep{2013ApJ...769...46A}. This object is just one of many examples
of ongoing destruction of larger bodies in our own solar system
\citep[see review of][]{2015arXiv150202361J}.

The solar debris disk could not yet be detected in surveys of disks
around other stars (e.g., with {\it Spitzer}, {\it Herschel} or {\it
WISE}).  Based on models of the solar system's dynamical evolution,
\cite{2009MNRAS.399..385B} predict its brightness variation over time. 
When surveys of debris disks reach comparable limits to the Kuiper
belt's dust fractional luminosity ($\sim 10^{-7}$), we will be better
able to judge whether many such faint systems exist.

Observations of individual planetesimals are not possible around other
stars; instead, we see the dust emission produced by the collisional
cascade that represents the steady state evolution of many debris disks
\citep{2007ApJ...663..365W,2008ARA&A..46..339W}. The observational evidence for
debris disks shows that their planetesimal belts are long-lived,
existing around stars of any age, and that they can be anomalously
bright at late times, suggesting that stochastic as well as steady-state
processes act within them.  Depending on the wavelength observed and
even the resolution of the imaged region, the dust distribution
can appear very different. For instance, millimeter/centimeter grains
are preferentially observed near their radius of formation, since they
do not migrate far before being ground down to smaller
sizes. Therefore, these larger grains are the best tracers of the
planetesimal belts in external debris disks because they indicate the location of active collision processes
\citep{2006ApJ...639.1153W}. In contrast, scattered light imaging
highlights the location of much smaller grains, which are easily
pushed around in the system, moving inward due to PR drag or stellar
wind drag or outward due to radiation pressure or stellar (particle)
winds.  Emission from warmer dust is also seen at infrared
wavelengths, typically highlighting the location of asteroidal or
terrestrial zones of warm dust as illustrated by Figure 1 of
\cite{2014prpl.conf..521M}.  Generally, the timescales for the
evolution of warm dust (as seen by {\it Spitzer} and {\it WISE}) are
much shorter than the cold dust detected in the Kuiper belt analogues
(by {\it IRAS}, {\it Spitzer} and {\it Herschel}), which explains why
warm dust less frequently detected, with incidence rates of a few
percent, rather than the 15-25\% for cold dust
\citep{2014prpl.conf..521M}.

Debris disks are considered signposts of planetary systems because
evidence of a bright debris disk indicates that the planetesimals in
the disk have been stirred, inducing collisions. For a recent review
of stirring mechanisms, see Section 6 of \cite{2014prpl.conf..521M}
and references therein. \cite{2012MNRAS.424.1206W} and
\cite{2014A&A...565A..15M} find evidence for a correlation between the
presence of a bright debris disk and the presence of low-mass planets,
many of which were not known when the systems were targeted by {\it
Herschel} surveys for debris disks.  In parallel, the {\it Herschel}
Survey 4 Kuiper Belts Around RV-detected Planet hosts (SKARPS)
targeted 97 planet host stars and also find a higher correlation of
brighter disks with low-mass planet systems (i.e., those in which the largest
planet mass is $<$ a Saturn mass) than with systems in which a higher
mass planet is present, though the number of low-mass planet systems
are still small (Bryden et al., private communication). The limited
number of low-mass planet systems prevents statistical analysis of the
joint DUNES/DEBRIS dataset from firmly establishing the correlation
\citep{2015ApJ...801..143M}, although we note that the statistics can
improve in the DUNES and DEBRIS samples as more planets are identified
in nearby systems.

\cite{2015Ap&SS.357..103W} recently authored a detailed summary of the 
steps that occur as a protoplanetary disk becomes a debris disk. We do
not attempt to replicate that work here.  In the first part of this
section, we briefly discuss recent results that provide insight into the
critical epoch of planetesimal formation based on observations. We
then discuss the implications of the detections of gas disks around
debris dust systems and the implications for interpretating measured
abundances in terms of the dynamical history of those systems. We then
step back from extrasolar debris disks to highlight what can be
learned about the origins of the solar system from its
planetesimal population. The observations of the solar system's
planetesimal population are interpreted based the reasonable
assumption that the chemical features observed in asteroids and comets
today were inherited from the local disk at the time and place of the
planetesimals' formation (i.e., the transition disk phase), and
the system subsequently evolved dynamically.  Finally, we highlight the
deuterium fraction as a key disciminator of the solar system's
planetesimal history and its potential in external debris disks.

\section{Tracing the critical epoch of planetesimal formation}
\label{ss:epoch}

Planetesimals, i.e., solid
bodies ranging in size from 10s - 1000s of km, may form very early in
protoplanetary disks.  The agglomeration process must overcome the
meter-size barrier to produce oligarchs capable of assembling into the
cores of giant planets or terrestrial planets in their own right
\citep[e.g., see][]{1980Icar...44..172W,2015A&A...579A..43C}. 
Observations of many planetary systems reveal that not all the
planetesimals are assembled into larger bodies, however. Many
asteroids and comets remain to undergo collisional evolution. Based on
recent {\it Herschel} and {\it Spitzer} surveys, $\sim 25$\% of A stars
\citep{2014MNRAS.445.2558T,2009ApJ...705..314S} show evidence of a 
measured far-infrared excess (to the survey depths) with a detectably
higher rate of detection for younger stars in the {\it Spitzer}
sample. {\it Herschel} surveys of solar type (FGK) stars, find a
detection rate of $\sim 20$\% \citep{2013A&A...555A..11E}, with
evidence of a significant decrease in the rate of measured incidence
from $\sim 25$\% for F-type stars to $\sim 15$\% for G/K-type stars
(Sibthorpe et al.~2016, in preparation).  The survey data do not
yet probe disks as faint as our own Kuiper Belt.  The range of system
ages where dust from planetesimal collisions can be observed is a
strong indication of the longevity of debris disks. 

In contrast to the debris disk phase, the period during which
the planetesimals can be assembled must be relatively short ($<$ 10
Myr) since the gaseous disk components are significantly depleted
for most systems after that time.  This timescale is also reinforced
by the planetesimal population of the solar system itself, as we
discuss below. To study the assembly of planetesimals, protoplanetary
and transition disks must be observed. 

With the advent of the Atacama Large Millimeter/submillimeter Array
(ALMA), thermal imaging of circumstellar disks with comparable
resolution to scattered light imaging and high sensitivity has at last
become feasible.  Even in its early science phase, ALMA has observed
many transition era disks, typified by the loss or strong depletion of
the inner disk material \citep[see the review
by][]{espaillat_etal_2014}; recent ALMA results establish that some
gas does remain the inner regions cleared of dust
\citep{2015A&A...579A.106V}.  Observations have revealed very
similar morphologies in several transition disks, including dust disks
that are typically much more radially confined than the gas disk
components, and asymmetries are also commonly observed \citep[e.g., HD 100546,][]{2014ApJ...788L..34P,2014ApJ...791L...6W}.  Most significantly, many disks exhibit highly asymmetric dust dust structures, suggesting that that planetesimal formation processes may be highly
localized in the disk. For example, 
\cite{2013Sci...340.1199V} observed a significant azimuthal asymmetry 
in the disk around Oph IRS 48, revealing an enhanced mm-grain ``dust
trap'' containing up to 9 Earth masses of material (assuming grains up
to 4 mm in size) while the 18.7 $\mu$m emission from VISIR reveal that
the smaller grains were much more uniformly distributed in the disk.
The authors suggest this dust trap feature is consistent with a
vortex, which confines material both radially and azimuthally.
\cite{2014ApJ...783L..13P} note strong azimuthally asymmetric
distributions of dust consistent with vorticity in the disks of SAO
206462 and SR21, and HD 142527
\citep{2013Natur.493..191C,2013PASJ...65L..14F} exhibits a disk in which dust is confined to a horseshoe pattern peaking at the northern edge of the disk. 

The composition of planets and planetesimals varies according to the
constituent materials available where they form in the protoplanetary
disk. Therefore, the variations in conditions across the disk have a significant impact on the larger bodies produced. For example, condensation fronts of various volatile species
form boundaries between different types of compositions \citep{2011ApJ...743L..16O}.  Icy grains
are more porous and grow more quickly than pure silicates. In
addition, many studies show that surface mass densities are enhanced
near snow lines, creating favourable conditions to form or grow
planetesimals.  \cite{2015ApJ...806L...7Z} illustrate that the
spectral index map derived from ALMA data toward the protoplanetary
disk HL Tau \citep{2015ApJ...808L...3A} reveals that the gaps observed
in the disk correspond well to the expected location of condensation
fronts of water ice, ammonia and hydrates formed from amorphous water
ice and other species. \cite{2015ApJ...806L...7Z} show that the flux
ratio between 1.3 mm and 0.87 mm is more extreme in the areas of lower
emission, and they suggest that the variation observed is best
explained by a model with two dust populations, one of which is the
product of signficant grain growth within the ``dips'' visible in the
disk.

The HL Tau data, and the theoretical models with which they are consistent,
suggest that this era of disk evolution may be the most critical in
the assembly of planetary masses. The assemblage of so much material
concentrated in small regions of the disk is the key to pebble
accretion scenarios theories \citep[see][]{2007Natur.448.1022J}.
Pebble accretion allows solids, through mass accretion over a short
period of time, to grow from relatively small ($< 1$ m) bodies to
planetary scale oligarghs or embryos, vaulting over the meter-size
barrier, which has been an impediment to core accretion theories of
planet formation.

\begin{figure}[tbh]
\includegraphics[trim=1cm 12cm 3cm 1cm,clip=true,width=0.8\textwidth]{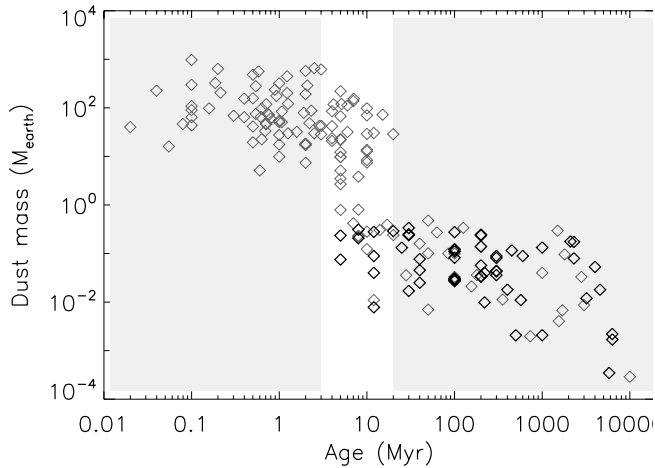}
\caption{Dust masses of circumstellar disks inferred from millimetre and submillimetre measurements. Data were compiled by \cite{panic2013} to which new data from the SCUBA-2 Observations of Nearby Stars (SONS) Survey have been added (darker symbols). Beyond 10 Myr, disks are
  significantly less massive (100-1000 $\times$) compared to younger,
  protoplanetary disk systems, representing a rapid decline in the
  mass located in mm-sized dust grains during the transition to the
  debris disk phase. There is an uncertainty to any age of at least a
  factor of 2, though the relative ages shown should be more
  accurate. The broadest range of masses are observed between $3--20$
  Myr, during the transition disk era.  We note that while typical
  distances of star-forming regions place detections of typical debris
  disk masses outside practical sensitivity limits, the dearth of high
  mass disks at late times is a real effect. We would be sensitive to
  very massive disks around nearby stars of 10+ Myr, and these are not
  seen. (image courtesy of Olja Pani\'{c} and the SONS Survey team.)}
\label{massdecline}
\end{figure}

Another window into the evolution of grains is offered by measurements
of dust emission in the submillimetre and millimetre regime. Because
the emission is optically thin and the opacity is relatively simply
modeled (as a power law in frequency), the flux density can be viewed
as a proxy for mass (in millimeter-sized grains), provided one has
some information about the dust temperature.  The dust masses of many
disks have been measured and can be compiled to gauge how dust mass
varies with age.  From observations at millimetre wavlengths, grains
larger than centimetre sizes will not be observable, so disks may
appear to diminish in mass as agglomeration occurs.  Figure
\ref{massdecline} shows that there is a marked decline of over two
orders of magnitude in the dust mass of millimetre-sized grains in
disks around the 10 Myr epoch, but the decline in fact extends over a
range from $3-20$ Myr.  This timescale for the decline of millimetre
continuum emission is longer than the 3 Myr timescale observed by
\cite{2001ApJ...553L.153H} during which time one half of the disks had disappeared in the infrared.  This makes sense because the infrared emission traces
emission closest to the star, whereas the millimeter emission
preferentially traces cooler grains further from the star, which take
longer to evolve or dissipate.

The fate of the millimetre grains lies in one of two paths: either
they are rapidly ground down to smaller dust grains, rendering them
susceptible to forces capable of rapidly removing the grains from the
system via radiation pressure or, for low mass stars, stellar wind
\citep[e.g.,][]{2006A&A...455..987A,2006ApJ...648..652S}. Either way,
their mass would be permanently removed from the disk, possibly
following a short period of anomalously bright emission at infrared
wavelengths and in scattered light due to the enhanced amount of small
grains in the system.  The alternative possibility is that the dust
grains undergo a period of rapid grain growth to sizes that do not
radiate efficiently at these observing wavelengths ($> 1$ cm),
rendering them invisible.

\section{The critical role of gas}
\label{ss:gas}

Due to the absence of gas detections in debris disks, they were
typically described as gas-poor, and the absence of gas emission was
considered characteristic of debris systems.  While this continues to
be a generally accurate paradigm, there are now almost a dozen systems
which have gas detections.  The earliest detections were of CI and CO
in UV absorption spectra toward $\beta$ Pictoris
\citep{2000ApJ...538..904R} and AU Microscopii via fluorescent H2 emission
lines \citep{2007ApJ...668.1174F}. More recently, detections are
typically measurements of CO in emission, e.g., $\beta$ Pictoris
\citep{2014Sci...343.1490D}, 49 Ceti
\citep{2005MNRAS.359..663D,2008ApJ...681..626H,2012ApJ...758...77Z,2013ApJ...771...69R,2014ApJ...796L..11R},
HD 21997
\citep{2013ApJ...776...77K}, HD 32297 (J.S.~Greaves et al. in
preparation), and several debris disk targets in the Sco-Cen
association (J.~Carpenter et al., in preparation). In the case of the
ALMA observations of $\beta$ Pictoris, the distribution of gas mirrors
that of the dust emission, with an extreme asymmetry to one side of
the edge-on disk, likely at the location of a recent collision
\citep{2014Sci...343.1490D}.  In several other disks, the CO
distribution is not ostensibly asymmetric and exhibits a Keplerian
rotation curve.  For example, the circumstellar disk of the 30 Myr old
star HD 21997 has been resolved by ALMA. Its dust distribution
is consistent with a debris disk \citep{2013ApJ...777L..25M}, while
the presence of the CO gas disk \citep{2013ApJ...776...77K} in such an
evolved source presents an ambiguity: is it remnant gas from a
protoplanetary disk in its last throes or is the gas, as well as the
dust, second generation, a product of the same collisions that produce
the dust?

\cite{2013ApJ...776...77K} and \cite{2012ApJ...758...77Z} before them have pointed out that the CO mass 
($4-8\times 10^{-2} M_\oplus$) estimated in the HD 21997 disk
indicates an unusually gas-rich disk {\it for a debris dust system},
suggesting HD 21997 \citep[and 49 Ceti,][]{2012ApJ...758...77Z} may
just represent atypically massive protoplanetary disks. If the
detected gas remains from the protoplanetary and transition disk
phase, then the CO is just a proxy for $H_2$ emission and the mass
could be as high as $26-60 M_\oplus$.  If this system is established
as a true hybrid system, in which the dust is generated by
collisional processing of larger bodies but the gas is not, then this
object may represent an example of the final phase of a transition
disk, before the remnant protoplanetary disk (small grains and
gas) is dissipated and the disk becomes a true debris disk
\citep{2015Ap&SS.357..103W}.

The possibility that gas removal is the final step in the
evolution of a protoplanetary disk to a debris disk is significant
because of the critical role played by gas in the final assembly of
planetesimals.  Gas acts to stabilize orbits of planetesimals and
planets, and in its presence, small dust grains are well coupled to
and co-orbiting with the gas disk. Slowing the relative velocities of
dust grains enhances agglomeration properties in the disk, as
described in the case of vortices in the previous section.  The gas is
also the primary reservoir of the magnetic field in the disk, which
can act to create dead zones where mass can be focussed to enhance
agglomeration processes \citep{2013ApJ...765..114D}.

In the solar system, the timing of gas dispersal can be constrained by the
size distribution of material being delivered from the Oort Cloud.
Small objects \citep[$<$ 10~km,][]{2007Icar..191..413B} 
would be more strongly affected than larger bodies by an increase in gas drag in the outer disk. If the delivery of
material into the Oort cloud occurs in the presence of significant
primordial gas, a reduced Oort Cloud results.  In our own solar system
this would be observable as a suppression in the number of 1-2 km size
Oort cloud comets.  Therefore, in the presence of gas, the material that would
normally have arrived in a system's Oort Cloud region remains in the
protoplanetary disk region, providing a substantial reservoir of
short-period comets \citep{2007Icar..191..413B}.  Significant gas
presence during the late stage re-ordering might then provide the
conditions needed for the substantial cometary number densities needed
to explain the significant collisional processing seen in some systems, i.e., CO levels which could require a cometary collision every 6 seconds \citep[e.g., 49 Ceti,][]{2012ApJ...758...77Z}.

As we will discuss below in more detail, the fact that gas chemistry
is highly temperature dependent for many processes means that there
will be variation in compositions in the gas throughout the disk, a
radial function of temperature, and even potentially scale height.
Over time, this, coupled with gas-solid chemistry, can lead to a
diversity of surface properties in the planetesimals which should
follow a systematic gradient as imparted from the disk. Once the gas
is removed from the system, those diverse properties serve as a record
of the location of those planetesimals in the disk, which should allow
us to determine the final location of those planetesimals before
significant dynamical evolution occurs. 

Most models of the solar system are in fact predicated on this
assumption, i.e., that the characteristics of the planetesimals we see
today reveal that the dynamical history of the solar system must
include several upheavals to account for the current observed
properties of the population.  The observed structure of the Kuiper
belt clearly reveals that Neptune at least migrated
\citep{1993Natur.365..819M} and mostly likely scattered from a
formation location near the present day locations of Saturn and Uranus
\citep{2003Natur.426..419L}.  Meanwhile, the presence of Main Belt
Comets and the variation among the surface properties of the various
asteroid classes may imply that Jupiter and Saturn also had a complex
migration history
\citep{2012M&PS...47.1941W}.  Such large scale restructuring events
may be observable as chemical gradient anomalies in exo-debris disks
when the gas is observable.

\section{Insights from the Solar System}
\label{ss:insights}

\begin{figure}[!htb]
 \centering
 \includegraphics[trim=2cm 5cm 2cm 6cm,clip=true,width=0.8\textwidth]{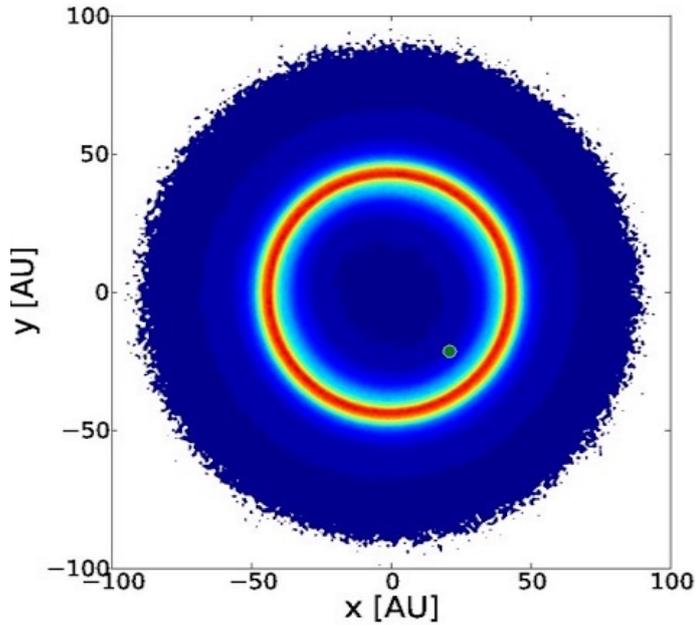}

 \caption{Number density contour plots of the debiased Kuiper belt looking down on the Solar System from above.  The position of Neptune is shown by a green circle. Classical objects dominate the number density, with resonant objects nearly invisible because of their much lower number density. Derived from \cite{2014IAUS..299..232L}.}
 \label{fig:xynumber}
\end{figure}

The debris disks of the solar system provide directly measurable
examples of the suspected sources of dust seen around other stars.
Observations of the current day structure of the Kuiper belt suggest
that the early Kuiper belt experienced a large scale
instability.  \cite{2009MNRAS.399..385B} provide a model of expected
emission during such a destabilizing event in our solar system and
compare that emission to detected debris disk systems.
\cite{2014IAUS..299..232L}  provides a model of the expected 
exo-debris disk emission that would be produced by the {\em current}
Kuiper belt (see Figure~\ref{fig:xynumber}).  The current day
fractional luminosity ($\sim 10^{-7}$) is many orders of magnitude
lower than the early disk predictions (a few $\times 10^{-3}$) from
\cite{2009MNRAS.399..385B}.  The orbital structure of the Kuiper belt
provides an excellent motivation for the interpretation of currently
observed exo-debris disks and their decay.

The solar zodiacal dust, which may be compared to warm dust around
other stars, is, in the majority, produced through disruptive
splitting of Jupiter family comets (JFCs) rather than through
sublimation or asteroid belt collisions \citep{nesvorny2010cometary}.
The size distribution of the zodiacal dust can be measured directly
and is found to be consistent with a collisional cascade that occurs
near the site of dust production (Jupiter's orbital distance) and then
transported inward through PR-drag.  Earlier in the evolution of our
planetary system we expect that the production of JFCs would greatly
exceed the current rate, due to a larger supply source in the region
of the Kuiper belt, resulting in an inner warm dust belt that would be
easily detected.  Thus, the detection of warm dust, so called
``exo-zodi'' debris disk components, may be an indicator of a large
cometary population that is undergoing disruption.  The
extrasolar disk systems in which warm dust has been detected exhibit
significantly stronger emission than is seen around the Sun; see
\cite{2013MNRAS.433.2334K} for a discussion of such systems and
their possible origins.  (As for cold Kuiper belt analogues, we cannot
yet detected warm excesses comparable to that of our own solar
system.)  The composition of the warm dust, having being produced
predominantly by objects that formed in the more distant
interplanetary region, is then not indicative of the local
planetesimal formation conditions, but rather those at large distances
where the temperatures, densities and chemical abundances would have
been very different.

The time scales for the process of accretion are well constrained by
measurements of isotopic ages of the meteorite population.  Ages
are ascertained by measurements of the decay products of $^{26}$Al ($^{26}$Mg) in comparison to lead-lead dating (i.e., comparisons of
$^{207}$Pb or $^{206}$Pb to $^{204}$Pb). Comparing ages of adjacent
components of a meterorite reveal the timescale for its creation,
because once $^{26}$Al is created, it must be bound up quickly before it undergoes decay. Applying these dating techniques
to calcium-aluminum-rich inclusions (CAIs) and other components
(called ``grains'') of H4 (among the most primordial and least heat-processed)
chondrites indicate that these ``grains'' formed and underwent
metamorphic processes within a few million years of the formation of
the CAIs \citep{zinner2002aluminum}.

This time scale for formation of $^{26}$Mg-rich grains sets a short
upper bound on the planetesimal formation timescale as the grains are
taken up into the chondrites prior to the decay of $^{26}$Al.  These rapid (few
million year) timescales are consistent with evidence that stars older
than about 10 Myr are depleted in gas and do not possess
protoplanetary dust disks (see Section~\ref{ss:gas}).  Models of the
processes of disk evolution and planetesimal accretion must keep these
formation timescales in mind.

The observable material in the asteroid and Kuiper belts have chemical
properties that hold the signature of the thermal histories of those
objects.  Within the asteroid belt there is a clear gradient in
reflected surface properties from refractory rich to volatile rich, while the Kuiper belt exhibits a diversity of processed surface
ices.  The gradients seen in the asteroid belt are nominally
reflective of the formation locations of those objects.  There are,
however, anomalous objects whose properties appear more consistent
with formation elsewhere within the disk, and these may indicate
post-formation transport of already formed objects
\citep{2012M&PS...47.1941W}.  If the Kuiper belt ices are indicative
of variations in formation location, then the diversity of surfaces
reflects a large scale re-ordering of the system which would have left
a detectable but very short-lived excess IR emission
\citep{2009MNRAS.399..385B}.  Although collisions within the belt are
currently rare, they were likely frequent during this re-ordering
processes.  As the transport of material was from the warmer (interior
to Neptune) zone to the colder zone, the gas liberated from the ice
during the collisional processing would be deficient in the volatile
species that would have survived had the colliding objects formed
in-situ.  Detection of abundance anomalies in second generation gas in
extrasolar debris disks may be a signature that such large scale
re-ordering has occurred in those systems.

The observed orbital diversity in the Kuiper belt requires the giant
planets to have experienced a large scale migration phase after their
formation.  Pluto's orbital resonance with Neptune has long been
recognized the strongest evidence of a past migration phase for
Neptune \citep{1993Natur.365..819M}.  Subsequent surveys of the Kuiper
belt,
see \cite{Bannister:o9a-KucX} for a complete listing,
have revealed a rich orbital resonance structure that is not well
matched by models that do not include Neptune migration
\citep{2012AJ....144...23G} while clearly indicating that smooth
migration cannot have been the complete story.  The Outer Solar
System Origins Survey \citep[OSSOS,][]{Bannister2015} 
is specifically designed to measure the relative abundances
of the various resonance populations and determine the migration
history of the outer solar system.  Within the non-resonant
populations, the cold component of the main classical Kuiper belt
\citep[see][for definition]{2009AJ....137.4917K} is composed of two
distinct subcomponents: the Kernel and the Stirred
\citep{2011AJ....142..131P}.  The Kernel component may be explained as
the result of a `skip' in Neptune's migration during a secular
rearrangement of the outer solar system orbital architecture
\citep{2015arXiv150606019N}.  The structural rearrangements would have
resulted in increased collisional cross-sections within sub-components
of the Kuiper belt, likely resulting in the dust-ring type structures
seen in extrasolar debris disks today.  

It is possible that the Kernel subcomponent of the Kuiper belt formed
in situ, rather than through migration processes from elsewhere in the
disk. Then, the inferred mass of the protoplanetary disk, particularly
in the region of the Kernel component, presents a challenge for
`standard' accretion scenarios.  The reason in situ formation
must be considered is that nearly 100\% of the larger members of the
cold classical Kuiper belt objects (KBOs) are binary systems
\citep{2014DPS....4650705N}, with a large fraction having the
separation between the two components being a substantial fraction of
the systems Hill radius.  These wide-binary cold classical KBOs are
very weakly bound and thus easily disrupted.  The weakness of the
binding argues in favor of their formation `in-situ' at their current
physical location \citep{2012ApJ...744..139P}.  Indeed, the systems
are so weakly bound that they would not survive in a bath of smaller
objects produced during a collisional cascade
\citep{2012ApJ...744..139P}.  {\em These two constraints appear to
require that the Kuiper belt region itself experienced in-situ
planetesimal accretion that must have proceeded in a low-density
environment where collisional disruption was minimal}.  Formation
within a dust-trap inside the protoplanetary disk may provide these
conditions.  Such dust-traps may have already been observed in
transition disk systems
\citep[i.e.,][]{2015ApJ...810L...7V,2015A&A...584A..16P}.  In such a
scenario, where there is no nearby massive planet, the planetesimal
disk would be left relatively undisturbed, post-formation in the trap,
resulting in little debris dust.  This suggests that such cold
accretion systems may build up, undetected.  The ``cold disks''
reported by \cite{2013ApJ...772...32K} are also suggestive of such a
process.

Unlike the cold classical components, the dynamically hot, high
inclination distribution component of the Kuiper belt appears to have
been implanted by a large scale scattering process. The hot component
appears to stretch from just beyond the orbit of Neptune out to 60+ AU
\citep{2009AJ....137.4917K}.  Additionally, the inclination distribution 
of the hot component appears to require that the post scattering
migration of Neptune was slow
\citep{2015arXiv150406021N}.  This component possesses very little in the
way of large separation binaries \citep{2010DPS....42.2305G}
suggesting a different dynamical history for this population compared
to the Kernel component.  The hot component has surface reflectance
\citep{1998Natur.392...49T, 2012ApJ...749...33F, 2015A&A...577A..35P}
and size distributions
\citep{2004AJ....128.1364B, 2010ApJ...722.1290F, 2010Icar..210..944F,
2011AJ....142..131P} that appear to be distinct from the cold
classical KBOs.  The scattering event that produced the hot component
would likely have resulted in significant collisional processing and
dust production, especially if the event occurs in the inner solar system.  Detection of sudden or very high levels of dust
production in other systems
\citep[e.g., such as the detection of large amounts of warm dust by][]{2014Sci...345.1032M,2013ApJ...778...12M} could be
evidence that such large scale scattering events are common, as
suggested by the solar system evolution models that contain them.

The chemical compositions of today's cometary population may provide
some insight into the past restructuring processes.  Observations of
fractionation ratios in deuterated ice species such as DHO and DCN
provide probes of the temperature at which the molecules condensed
onto ice-forming grains \citep{2000Icar..148..513M}.  The ice
fractions and types of ice available provide additional constraints on
the formation conditions \citep{1988Icar...76..225A}.  Disentangling
which signatures result from interstellar origins
\citep{1996Sci...272.1310M} and which might be a signature of
re-arrangement of the planetesimal disk \citep{2011ApJ...734L..30K}
presents a significant challenge.

\subsection{Deuterium fractionation in the early solar disk}

One of the most distinctive imprints of the early disk is expected to
be the fractionation of deuterium, or the ``D/H ratio.''  In the low
temperatures of the protostellar core of a forming solar system,
fractionation of deuterium is expected to be enhanced in the gas phase
and locked into microscopic ice particles even before the disk is
formed \citep[e.g.][]{2011ApJ...734L..30K,2010ApJ...725L.172J}.  
The level of fractionation of water, defined as $f_{H_2O} =
\frac{[DHO]/[H_2O]}{[DH]/[H_2]}$, is enhanced to about 35 times solar,
while the level of fractionation of HCN, $f_{HCN} =
\frac{[DCN]/[HCN]}{[DH]/[H_2]}$, is over 150 times solar. When the
deuterium-enriched material is carried into the disk, the grains are
heated and the molecules are released into the gas phase.  What
fraction of the deuterated species relative to the hydrogen species
enters the gas phase depends on the individual molecule and the period
of heating. The closer they are to the forming star, the warmer the
grains become and more deuterium is released, and the fractionation
can then vary based on the local gas chemistry. In the most distant
parts of the disk, the grains may never get warm enough to release a
given species into the gas phase (e.g., 30 K for water or HCN), in
which case, the high ratio from the protostellar core should be
retained.  We therefore expect the D/H ratio to be large in the outer
disk but drop closer to the star where temperatures are higher and
deuterium enrichment is not favored.

Earth's value of $f_{H_2O}$ is consistent with the lower values
measured for primitive chondrites in the asteroid belt, an enrichment
factor about 4 times solar \citep[see, for example, Figure 3
of][]{2015Sci...347A.387A}.  Oort cloud comets, Saturn's moon
Enceladus and the Rosetta measurement of comet 67P/CG show much higher
values. The giant planets and the Sun have minimal enrichment factors
consistent with a homogeneous protosolar nebula. Chondrites in fact
show a large variation in $f_{H_2O}$, ranging from 4 to as high as 30,
suggesting some condensed from the gas of the solar nebula, while
others are consistent with formation in situ near the Earth.  Several
comets also show values consistent with Earth, but most show values
between 10-20 with large error bars (i.e., at least 2x
Earth). Interestingly, among the Jupiter family comets (JFCs), the
recent result of
\citep{2015Sci...347A.387A} from Rosetta shows that comet 67P/CG has a
very elevated value of $f_{H_2O}$, which is at odds with the other
members of its class.  The authors argue based on this
result that the JFCs may be highly hetergeneous, reflecting diverse
origins. Furthermore, the new measurement supports models that argue
in favour of an asteroidal origin for Earth's oceans and its
atmosphere, rather than a cometary one.

Interpretation of the variation in cometary D/H measurements in terms
of the D/H ratio in the solar disk is greatly complicated by the fact
that the objects observed have undergone dynamical evolution.  Their
D/H ratios were ``locked in'' at the values present in the gas phase
when the ices condensed onto the asteroids and comets we observe
today. We can't turn back the clock on the solar system, but we can
observe young systems at this critical phase to measure how their D/H
ratio varies.  Since the D/H ratio in the gas phase should exhibit
temperature (and hence radial) dependence, we should be able to
observe a radial dependence if we can resolve such a disk.

Models, like \cite{2011ApJ...734L..30K}, attempt to map the formation
location of cometesimals using the observed variation of $f_{H_2O}$ in
cometary bodies.  If it can be established that the D/H ratio of
a protoplanetary system had a temperature dependence, then the formation
distance of a given comet or asteroid can be recovered based on its
relative D/H ratio. Such a tool would allow the original configuration
of our solar system to be retraced, and importantly, untangle its
dynamical evolution.

\section{Tracing the chemical history of planetesimals}
\label{ss:examples}

Where did Earth's water and atmosphere originate? This has been one of
the key issues motivating the study of small bodies, comets and
asteroids, in the solar system. 

One of the key assumptions in the current model of the Solar system is
that periods of dynamical instability occurred in the early stages of
its evolution, particularly among the giant planets.  This dynamical
evolution was critical to Earth's habitability since these periods of
upheaval were likely responsible for the delivery of the water and
atmosphere we currently enjoy. Recent evidence, however, has called
into question the long-accepted cometary origins for terrestrial
oceans. Instead, it seems increasingly apparent that Earth's water has
its likely origin in asteroids and bodies which formed much closer to
the Earth
\citep{2015Sci...347A.387A,2014Icar..239...74O,2014ApJ...784...39A}.
This interpretation is based on the idea that chemical evolution must
also have occurred in the early solar system.  As planetesimals form
inside a protoplanetary disk, the chemical signatures and abundances
in the disk should be recorded in those solid bodies, creating a
record of the chemical diversity of the nascent solar disk.

What can this tell us about planet formation? 

The position of objects in the solar system has evolved dynamically
over time.  Interpretation of the solar system's evolution relies on
an assumption about the gas chemistry in the solar disk at the time
the planets and planetesimals formed.  These assumptions can be tested
by observing protoplanetary disks at late stages and debris disks at
early stages to directly measure the radial dependence of certain
chemical tracers that are believed to have varied significantly at the
time the planetesimal formation was completed.  The probability is
high, and it may seem obvious, that a variation in the D/H ratio was
present in the solar system's disk when the planetesimals were formed,
but this ratio has never been resolved in a disk of this era.
\cite{2012ApJ...749..162O} and \cite{2008ApJ...681.1396Q} measured 
the DCN abundances
in the 8-12 Myr old TW Hydra disk and report a global $f_{HCN}$ of
$1.7 \times 10^{-2}$, many times the expected value in the early solar
nebula, $(2.5 \pm 0.5) \times 10^{-5}$ \citep{2000Icar..148..513M},
suggesting this ratio should be detectable in other disks.

The most ideal targets would be those disks now being identified which
appear to have highly evolved dust properties, but retain gas
distributions that are potentially the remnants of a protoplanetary
disk, e.g., HD 21997 \citep{2013ApJ...776...77K,2013ApJ...777L..25M}.
Changes to the dust grains can occur independently from the gas, since
large pebbles are so poorly coupled to the gas that their
disappearance would not alter the gas disk architecture. This scenario
describes the situation in HD 21997 very well.  Therefore it is likely
that young debris systems exist in which the deuteration fraction has
been imprinted onto the planetesimal population, according to the
thermal chemistry at their formation location, and the gas phase is
still present to measure that D/H ratio as well.

\section{Summary}
\label{ss:concbirth}

The study of debris disks provides direct evidence of planetesimal
formation in other systems. Since debris disks must be stirred to see
the bright collisionally produced dust detected thus far around
20-25\% of AFGK stars, it can be presumed that in many, if not all,
cases, the debris disks are also signposts of planetary systems as
well. The timescales by which systems have evolved to exhibit debris
disk qualities puts strong contraints on the limits of time available
for giant planet formation; in actuality, the transition period from
protoplanetary levels of dust and gas to debris levels of dust (and
rarely detectable gas) is very abrupt.  For this
reason, young associations of known age are particular targets of
study. Some, such as the TW Hydra moving group ($\sim 8$ Myr) contain
disks that are protoplanetary around some stars and still capable of forming planets \citep[i.e., such as TW Hydra
itself,][]{2012ApJ...744..162A,2013Natur.493..644B} while other stars are already debris disk hosts
\citep[e.g., HR 4796A and TWA 7,][]{1998ApJ...503L..83K,2007ApJ...663.1103M}.

The detection of gas associated with young debris disks (and of course
late transition disks) has the potential to provide substantial
information about the dynamical evolution of systems outside the solar
system and provide important constraints on the evolution of the solar
system itself. Systems in the transition epoch between protoplanetary
and debris levels of mm-grain emission are ripe for detection of gas
disk components.  Therefore, observations of disks around the 10 Myr
epoch transition period may capture the last phase of coalescence of
solid material onto planetesimals, and therefore any gas detections
should reflect the abundances and compositions deposited onto the
planetesimals. Disks emerging from the transition as debris disks may
exhibit detectable secondary gas disks, in which observed gas, like
the dust, is a collisional product of the planetesimals.

The increased sensitivity and resolution of ALMA in particular is
enabling the detection of secondary gas disks and even gradients in
those disks, as evidenced by the recent detection of CO gas associated
with several debris disks in Sco-Cen (J.Carpenter et al.,~in
preparation) and the detection of the secondary gas disk asymmetry of
$\beta$ Pictoris \citep{2014Sci...343.1490D}.  To current sensitivity
limits, $\sim 20-25$\% of nearby stars (less for G/K stars
than A/F stars) show evidence of infrared excess emission associated
with dust emission of debris disks. Despite the increased number of
detected debris disks, the stronger dust and gas emission toward the
shorter phase transition disks suggest that they can reveal more
readily the location of planetesimal belt formation. This is
particularly true at ALMA's observing wavelengths and resolution, as
evidence by the strong body of work on sites of dust agglomeration
already in the literature.

\cite{2015Ap&SS.357..103W} have identified the
rapid growth of planetesimals prior to the loss of gas as one of the
critical phases of the birth of a debris disk.  This rapid formation
phase is also supported by observations of radio-nucleotide signatures
in the solar system.  The radial thermal gradient in a transition disk
should result in a radial variation of the refractory and volatile
material being taken up into planetesimals and in predictable
variations of thermally sensitive molecules such as DHO, DCN and
other deuterated species. Resolved radial variations in these species will
soon provide a detailed mapping of the thermal processes in the
remotely observed systems.

Detailed studies of dynamical and physical properties of Kuiper Belt
Objects are untangling the past evoluton of giant planets, a necessary
step in the understanding of giant planet formation in our solar
system.  Evidence of large scale re-ordering in the solar system
suggests that chemical signatures in secondary gas and dust grain
reflected light do not reflect the local conditions but provide
evidence of transport of material in the system.  Similarly, outside
the solar system, detailed studies of resolved dust and gas
distributions combined with known or future detections of planetary
perturbers could dissect the history of the planetary system's
dynamical evolution, since species trapped in planetesimal ices and
released by collisions provide a record of the location of formation.
This in effect would permit us to use the disk's ongoing collision
evolution to trace its dynamical history.

\begin{acknowledgements}
The authors acknowledge the efforts of O.~Pani\'{c} and W.~Holland, who provided the revised SONS data of Figure 1 prior to publication and S.~Lawler for a revision of Figure 2 for this publication. The authors also acknowledge the efforst of two referees, whose queries and suggestions improved this work.
\end{acknowledgements}

\bibliographystyle{aps-nameyear}      
\bibliography{issibirth}                

\end{document}